\documentclass[12pt]{article}
\usepackage{amssymb}
\usepackage{amsmath}
\usepackage{color}
\usepackage{verbatim}

\allowdisplaybreaks

\newtheorem{theorem}{Theorem}[section]

\newtheorem{conjecture}[theorem]{Conjecture}

\newenvironment{definition}[1][Definition]{\begin{trivlist}
\item[\hskip \labelsep {\bfseries #1}]}{\end{trivlist}}

\begin{document}
\title{Nahm's conjecture and coset models: a systematic search for matching parameters}
\author{S Keegan and W Nahm\\
\small{Dublin Institute for Advanced Studies, 10 Burlington Road, Dublin 4, Ireland.}\\
\small{skeegan@stp.dias.ie}}

\maketitle

\section{Introduction}
An interesting open problem in mathematics is the question of when a $q$-hypergeometric series is modular. This question is far from being solved completely, but a useful first step is to consider the following problem. Let $A$ be a positive-definite symmetric $r\times r$ matrix, $B$ a vector of length $r$, $C$ a scalar, each rational, and define
\begin{equation}
f_{A,B,C}(q)=\sum_{n=\left(n_1,\ldots,n_r\right)\in\left(\mathbb{Z}_{\geq0}\right)^r} \frac{q^{\frac{1}{2}n^tAn+B^tn+C}}{\left(q\right)_{n_1}\ldots\left(q\right)_{n_r}}, \label{fABC}
\end{equation}
where $(q)_n=\prod_{j=1}^n(1-q^j)$. It is convenient to note that $f_{A,B,C}(q)=q^C f_{A,B,0}(q)$. This series converges for $|q|<1$. Problem: describe the set of such $A$, $B$, $C$ for which~(\ref{fABC}) is a modular function. For $r=1$ this question has been answered completely by Don Zagier~\cite{DZ}. A conjecture suggested by one of us~\cite{DZ,WN} attempts to partially answer this question by suggesting a condition that the matrix $A$ must satisfy in order to guarantee the existence of $B$ and $C$ that lead to~(\ref{fABC}) being modular. A brief summary of progress in this direction, as well as a description of the conjecture itself, is given in~\ref{Nahm}. The conjecture, its background, and related topics are discussed in detail in~\cite{DZ,WN} from the points of view of physics and mathematics respectively\footnotemark.

\footnotetext{One of us (W.N.) would like to apologise for the title of this paper, but it might be too confusing to use a more oblique name.}

In many cases Nahm's conjecture correctly predicts values of $A$ which give rise to modular $f_{A,B,C}$. However, even in such cases, there is no simple way to calculate the corresponding values of $B$. The conjecture claims such values exist but gives no indication of how to compute them. Of course one could search systematically through numerous potential $B$-values, each time computing the corresponding function $f_{A,B,C}$ and checking modularity, but this is a slow process. It would be useful to have an algorithm that computes suitable $B$-values for a given matrix $A$. This paper takes a first small step in this direction.

In the context of modularity, series of the form~(\ref{fABC}) first arose as characters of rational conformal field theories (RCFTs), see~\cite{NRT}. To explain characters of this form, the corresponding RCFTs must be understood as limits of more general integrable quantum field theories (QFTs), see for example~\cite{WN}. These are integrable massive perturbations of CFTs, no longer themselves conformally invariant. Integrable QFTs can be studied by means of the thermodynamic Bethe ansatz (TBA), see for example~\cite{KM} and references therein. For the current status of research see~\cite{LZ} and its references. Using the TBA approach, information can be extracted from a massive integrable QFT once its scattering matrix is known. This includes the values of $B$ and $C$. The matrix $A$ of~(\ref{fABC}) is essentially the phase difference of the integrable QFT $S$-matrix between large positive and large negative rapidity, while the equations~(\ref{xeqns}) of Nahm's conjecture (see~\ref{Nahm}) are a limiting case of the TBA equations. If $A$ comes from an integrable deformation of a CFT, one could potentially find the characters (and appropriate values of $B$ and $C$) by reconstructing the scattering matrix from $A$, but little is known about the feasibility of this approach for general $A$.

While this will be an interesting approach for future study, the current paper is concerned with a simple comparison between RCFT characters and $q$-series of the form~(\ref{fABC}). Approaching the problem of modular $q$-series from the point of view of conformal field theory puts a wealth of tools at our disposal, most importantly reliable formulae for calculating CFT characters. Formulae for calculating minimal model characters are given in~\cite{diF}, while the technique for calculating coset characters is described in~\cite{diF,VK}. Many choices of the matrix $A$ correspond to particular models in conformal field theory, and where such models can be successfully identified their characters should be closely related to the series~(\ref{fABC}). For such choices of $A$, values of $B$ and $C$ can be chosen and the series~(\ref{fABC}) compared directly to characters of the corresponding CFT model. This is a fairly straightforward process as it involves numerically checking equality of $q$-series (up to a certain order), rather than the tedious process of checking modularity. For a particular choice of $B$ and $C$, if~(\ref{fABC}) is equal to any linear combination of CFT characters, then it is necessarily modular, and we have thus succeeded in finding suitable values of $B$ and $C$. This paper studies two particular families of matrices $A$, both chosen for their close links to conformal field theory. For each family, a numerical study of a number of examples establishes a pattern for the $B$-values. While our results are not an algorithm for computing $B$ given a general matrix $A$, they do suggest that a more comprehensive search for such an algorithm may be worthwhile.

The paper is laid out as follows. Section~\ref{ACFT} describes the families of matrices $A$ and their links to CFT. Section~\ref{search} discusses the method used to search for $B$-values. Sections~\ref{mm} and~\ref{coset} contain the results of these searches, including general formulae for calculating $B$-values. A number of interesting phenomena arise during these calculations and are mentioned briefly in section~\ref{coset}. Nahm's conjecture, while not central to this paper, is certainly relevant, and is stated in~\ref{Nahm}. \ref{appAsym} contains asymptotic formulae used in the search for $B$-values.

\section{The relationship between the matrix $A$ and conformal field theory} \label{ACFT}
\subsection{Pairs of Dynkin diagrams and CFT} \label{Dynkin}
Certain integrable models can be described by pairs $(X,Y)$ of ADET Dynkin diagrams\footnotemark.  Such models have equations of the form $x=\left(1-x\right)^A$, where $A = \mathcal{C}(X)\otimes \mathcal{C}(Y)^{-1}$, $\mathcal{C}$ denotes a Cartan matrix, and $\otimes$ is the Kronecker product of two matrices. The effective central charge is known or conjectured to be
\begin{equation}
c_{eff}(X,Y)=\frac{r(X)r(Y)h(X)}{h(X)+h(Y)}, \label{ceff}
\end{equation}
where $r$ denotes the rank, and $h$ the dual Coxeter number of a Lie algebra. See~\cite{GT} and references therein. The pairs $(A_1,T_n)$ and $(A_1,A_n)$ are studied in this paper, with corresponding matrices given by $A=\mathcal{C}(A_1)\otimes \mathcal{C}(T_n)^{-1}$ and $A=\mathcal{C}(A_1)\otimes \mathcal{C}(A_n)^{-1}$ respectively. These particular families are chosen because they can be identified with known CFTs, see sections~\ref{Amm} and~\ref{Acoset}. Furthermore, matrices $A$ of this form are among the easiest to deal with computationally because of the one-dimensional factor $\mathcal{C}(A_1)$.

\footnotetext{ADE Dynkin diagrams correspond to the classical simple Lie algebras $A_n = \frak{su}(n+1)$ and $D_n = \frak{so}(2n)$, and the exceptional Lie algebras $E_6$, $E_7$ and $E_8$. The `tadpole' diagram $T_r$ is got by folding the diagram $A_{2r}$ in the middle to get a pairwise identification of vertices $T_r = A_{2r}/\mathbb{Z}_2$. The Cartan matrix of $T_r$ is identical to that of $A_r$ except for the entry $\mathcal{C}(T_r)_{rr} = 1$. Its dual Coxeter number is $h(T_r) = 2r+1$.}

\subsection{Minimal models} \label{Amm}
By equation~(\ref{ceff}), the model $(A_1,T_n)$ has effective central charge
$$c_{eff}(A_1,T_n)=1-\frac{3}{2n+3}.$$
The $(p,2)$--minimal model has effective central charge
$$c_{eff}=1-\frac{3}{p}.$$
Choosing $p=2n+3$ allows the $(2n+3,2)$--minimal model to be identified with the model described by the pairs of Dynkin diagrams $(A_1,T_n)$. Some aspects of this family were studied previously in~\cite{NRT}. For more information on minimal models see, for example,~\cite{diF}.\\

\subsection{Coset models} \label{Acoset}
By equation~(\ref{ceff}), the model $(A_1,A_n)$ has effective central charge
$$c_{eff}(A_1,A_n)=\frac{2n}{n+3}.$$
The coset model $\widehat{su}(2)_{n+1}/\widehat{u}(1)$ has central charge
$$c=\frac{3(n+1)}{n+3}-1=\frac{2n}{n+3}.$$
This equality of central charges is good evidence that, for unitary models, the model described by the pair of Dynkin diagrams $(A_1,A_n)$ is exactly the coset model $\widehat{su}(2)_{n+1}/\widehat{u}(1).$ For more details on coset models see, for example,~\cite{diF,GKO}.

\section{The search for suitable $B$-values} \label{search}
The following method is used to search for $B$-values:
\begin{enumerate}
\item{{\it Choose a matrix $A$.}\\
In this paper we consider only those matrices described in section~\ref{Dynkin}.}
\item{{\it Identify the corresponding conformal field theory.}\\
As described in sections~\ref{Amm} and~\ref{Acoset}.}
\item{{\it Calculate the CFT characters.}\\
A thorough description of the methods used to calculate CFT characters can be found in~\cite{diF}. To save space we don't repeat the description here as the methods are standard and comprehensive overviews are available in the literature.}
\item{{\it Choose a range over which to search for $B$-values.}\\
To start we assume that each component of $B$ satisfies $b_i\in\mathbb{Z}$, and search over the range $-8\leq b_i\leq 8$. As necessary, the search is extended to include $b_i\in\mathbb{Z}/2$, $b_i\in\mathbb{Z}/3$, $b_i\in\mathbb{Z}/4$,\ldots. The choice of denominator usually changes as the choice of $A$ changes. (As the rank of $A$, and hence the length of $B$, increased it was necessary to restrict the scope of the search to $-2\leq b_i \leq 2$ for computational reasons, however for the examples studied this range still allowed us to find appropriate $B$-values.) The search continued until no further $B$-values could be found.}
\item{{\it Use asymptotic formulae to immediately eliminate many of the potential $B$-values.}\\
Useful information can be obtained by studying the asymptotic behaviour of  $f_{A,B,C}(q)$ as $q\rightarrow 1$. For example Don Zagier~\cite{DZ} used asymptotics to prove Nahm's conjecture for rank $1$, although unfortunately this method is too computationally intensive to be usefully applied to larger rank at present~\cite{MV}. In the present paper two asymptotic formulae are used to quickly exclude many potential $B$-values. Both formulae are stated in~\ref{appAsym}, but for a detailed description of their origins see~\cite{DZ,MT,MT1}. The first formula~(\ref{asymptotic1}) originated in the PhD thesis of Michael Terhoeven~\cite{MT} and calculates $C$ given $B$. The second formula~(\ref{asymptotic2}) doesn't seem to be stated explicitly in the literature but was calculated by the authors of this paper using exactly the methods described in~\cite{MT,MT1}. Both formulae are used to speed up the search for $B$-values as follows: for each $B$-value under consideration, check firstly that it satisfies equation~(\ref{asymptotic2}), and secondly that it gives rise to a rational value of $C$ using equation~(\ref{asymptotic1}). Any $B$-value that doesn't satisfy these conditions can be immediately excluded. This quickly rules out a large number of unsuitable $B$-values.}
\item{{\it For remaining $B$-values calculate $f_{A,B,C}$ and compare to known CFT characters.}\\
Use~(\ref{fABC}) to calculate $f_{A,B,C}$ (with $C$ calculated using~(\ref{asymptotic1})). This series is compared numerically (in most cases to order $20$) to known CFT characters. If $f_{A,B,C}$ is equal to a linear combination of CFT characters for any value of $B$ then we have successfully identified a $B$-value that leads to modular $f_{A,B,C}$.}
\end{enumerate}

\section{The family $A=\mathcal{C}(A_1)\otimes\mathcal{C}(T_n)^{-1}$ and the $(2n+3,2)$--minimal model} \label{mm}
This section examines matrices of the form $A=\mathcal{C}(A_1)\otimes\mathcal{C}(T_n)^{-1}$. The corresponding CFT is the $(2n+3,2)$--minimal model, see section~\ref{Amm}. For $n=1,2,3,4\ldots$ we calculate the appropriate $B$-values using the method of section~(\ref{search}), until a general pattern emerges for the $B$-values. One example and the general pattern are reported below.

\subsection{Example - the $(5,2)$--minimal model}
$A=\mathcal{C}(A_1)\otimes\mathcal{C}(T_1)^{-1}=2$, $B=(b_1)\in\mathbb{Q}$, $C\in\mathbb{Q}$, and the $q$-series~(\ref{fABC}) is given by
$$f_{A,B,C}=q^C\sum_{n_1=0}^{\infty}\frac{q^{n_1^2+b_1n_1}}{(q)_{n_1}}.$$
The $(5,2)$--minimal model has two distinct irreducible characters:
\begin{eqnarray*}
\chi_{1,1}^{(5,2)} &=& q^{\frac{11}{60}}\left(1+q^2+q^3+q^4+q^5+2q^6+\ldots\right),\\
\chi_{1,2}^{(5,2)} &=& q^{-\frac{1}{60}}\left(1+q+q^2+q^3+2q^4+2q^5+3q^6+\ldots\right).
\end{eqnarray*}
A search for $B$-values (with $-5 \leq b_1 \leq 5$) gives the following results:
$$
\begin{array}{llcl}
C=-\frac{1}{60} & \quad B=(0) & \Rightarrow & f_{A,B,C}=\chi_{1,2}^{(5,2)},\\
\\
C=\frac{11}{60} & \quad B=(1) & \Rightarrow & f_{A,B,C}=\chi_{1,1}^{(5,2)}.
\end{array}
$$

\subsection{A general pattern of $B$-values}
For the matrix $A=\mathcal{C}(A_1)\otimes\mathcal{C}(T_n)^{-1}$, a general formula emerges for $B$-values that lead to modular $f_{A,B,C}$. There are $n+1$ such values, each of length $n$:
\begin{eqnarray*}
B &=& \left(0,0,0,\ldots,0,0,0\right),\\
B &=& \left(0,0,0,\ldots,0,0,1\right),\\
B &=& \left(0,0,0,\ldots,0,1,2\right),\\
B &=& \left(0,0,0,\ldots,1,2,3\right),\\
&\vdots&\\
B &=& \left(0,0,1,\ldots,n-4,n-3,n-2\right),\\
B &=& \left(0,1,2,\ldots,n-3,n-2,n-1\right),\\
B &=& \left(1,2,3,\ldots,n-2,n-1,n\right).
\end{eqnarray*}

\section{The family $A=\mathcal{C}(A_1)\otimes\mathcal{C}(A_{k-1})^{-1}$ and the coset $\widehat{su}(2)_k/\widehat{u}(1)$} \label{coset}
\subsection{Coset characters and their relation to the functions $f_{A,B,C}$} \label{cosetfABC}
The aim of this section is to investigate exactly how coset characters are related to the $q$-series $f_{A,B,C}$ of~(\ref{fABC}). Comments are made based on patterns that emerged from the study of members of the family $A=\mathcal{C}(A_1)\otimes\mathcal{C}(A_{k-1})^{-1}$. It appears that the $f_{A,B,C}$ are equal to sums of coset characters, and that suitable sums of characters can be predicted from the equation
\begin{equation}
\chi_l^{(k)}(q,u)=\sum_{\substack{m=-k+1\\m+l\equiv0\mod 2}}^{k}\chi_{\{l;m\}}(q)K_m^{(k)}(q,u), \label{decomposition}
\end{equation}
where $\chi_l^{(k)}$ denotes an $\widehat{su}(2)_k$ character, $K_m^{(k)}$ denotes a $\widehat{u}(1)$ character, $\chi_{\{l;m\}}$ is a character of the coset $\widehat{su}(2)_k/\widehat{u}(1)$, and $u$ is a variable independent of $q$. (\ref{decomposition}) is a set of $k+1$ equations describing the decomposition of $\widehat{su}(2)_k$ characters into sums of products of a coset character and a $\widehat{u}(1)$ character. See~\cite{diF} for more details. As a result of field identification, see~\cite{diF}, half of these equations are redundant (don't produce any new coset characters), therefore it is only necessary to study the equations for the $\widehat{su}(2)_k$ characters $\chi_0^{(k)},\ldots,\chi_{\frac{k}{2}}^{(k)}$ when $k$ is even, and the equations for the $\widehat{su}(2)_k$ characters $\chi_0^{(k)},\ldots,\chi_{\frac{k+1}{2}}^{(k)}$ when $k$ is odd. This can perhaps be seen more clearly in the examples of section~\ref{SetOfB}. For the $12$ examples studied, values of $B$ and $C$ have been found that make $f_{A,B,C}$ equal to the sum of coset characters
\begin{equation}
\sum_{\substack{m=-k+1\\ m+l\equiv0\mod 2}}^{k}\chi_{\{l;m\}}, \label{important}
\end{equation}
for each $l=0,1,2,\ldots,\frac{k}{2}$ when $k$ is even, and for each $l=0,1,\ldots,\frac{k+1}{2}$ when $k$ is odd. We are curious whether this is true for all matrices $A$ whose corresponding CFT is a coset model. No similar statement has been found in the literature. If~(\ref{important}) turns out to be a method for relating the series $f_{A,B,C}$ to coset characters in more general cases it will be important to understand its role in conformal field theory. This will be an interesting project for the future.

\subsection{Calculation of $B$-values -- Examples \label{SetOfB}}
{\it Example: The coset $\widehat{su}(2)_2/\widehat{u}(1)$\\}
The coset $\widehat{su}(2)_2/\widehat{u}(1)$ is the model described by the pair of Dynkin diagrams $\left(A_1,A_1\right)$. The corresponding matrix is $A = \mathcal{C}(A_1)\otimes \mathcal{C}(A_1)^{-1} = (2) \otimes \left(1/2\right)=(1)$. $B=\left(b_1\right)\in\mathbb{Q}$, $C\in\mathbb{Q}$, and the function $f_{A,B,C}$ is given by
$$f_{A,B,C}=q^C\sum_{n_1=0}^\infty\frac{q^{\frac{1}{2}{n_1}^2+b_1n_1}}{\left(q\right)_{n_1}}.$$
For this coset model the decomposition~(\ref{decomposition}) consists of the three equations
\begin{eqnarray*}
\chi_0^{(2)} &=& \chi_{\{0;0\}}K_0^{(2)} + \chi_{\{0;2\}}K_2^{(2)},\\
\chi_1^{(2)} &=& \chi_{\{1;-1\}}K_{-1}^{(2)} + \chi_{\{1;1\}}K_1^{(2)},\\
\chi_2^{(2)} &=& \chi_{\{2;0\}}K_0^{(2)} + \chi_{\{2;2\}}K_2^{(2)},
\end{eqnarray*}
so that by~(\ref{important}) we expect to find $f_{A,B,C}$ equal to the sums of coset characters\footnotemark
$$\chi_{\{0;0\}}+\chi_{\{0;2\}} \quad \qquad \mbox{or} \quad \qquad 2\chi_{\{1;1\}},$$
for some choices of $B$. The search of section~\ref{search} gives the following results:
$$
\begin{array}{llcl}
C=-\frac{1}{48} & \quad B=(0) & \Rightarrow & f_{A,B,C}=\chi_{\{0;0\}}+\chi_{\{0;2\}},\\
\\
C=\frac{1}{24} & \quad B=\left(-\frac{1}{2}\right) & \Rightarrow & f_{A,B,C}=2\chi_{\{1;1\}}.\\
\end{array}
$$

\footnotetext{Note that field identification results in only three distinct coset characters:
\begin{eqnarray*}
\chi_{\{0;0\}} &=& q^{-\frac{1}{48}}\left(1+q^2+q^3+2q^4+2q^5+3q^6+3q^7+5q^8+5q^9+7q^{10}+\ldots\right), \\
\chi_{\{0;2\}} &=& q^{\frac{23}{48}}\left(1+q+q^2+q^3+2q^4+2q^5+3q^6+4q^7+5q^8+6q^9+\ldots\right), \\
\chi_{\{1;1\}} &=& q^{\frac{1}{24}}\left(1+q+q^2+2q^3+2q^4+3q^5+4q^6+5q^7+6q^8+8q^9+\ldots\right).
\end{eqnarray*}}

\vspace{5mm}

\noindent
{\it Example: The coset $\widehat{su}(2)_4/\widehat{u}(1)$\\} \label{su24}
The matrix $A$ corresponding to this coset is
$$A = \mathcal{C}(A_1)\otimes \mathcal{C}(A_3)^{-1} = (2) \otimes \left( \begin{array}{ccc} 3/4 & 1/2 & 1/4\\ 1/2 & 1 & 1/2\\ 1/4 & 1/2 & 3/4 \end{array} \right)
= \left( \begin{array}{ccc} 3/2 & 1 & 1/2\\ 1 & 2 & 1\\ 1/2 & 1 & 3/2 \end{array} \right).$$
By~(\ref{important}), we expect to find $f_{A,B,C}$ equal to the linear combinations of coset characters
$$\chi_{\{0;0\}}+2\chi_{\{0;2\}}+ \chi_{\{0;4\}} \qquad \mbox{or} \qquad 2\chi_{\{1;1\}}+2\chi_{\{1;3\}} \qquad \mbox{or} \qquad 2\chi_{\{2;0\}} + 2\chi_{\{2;2\}},$$
for certain choices of $B$. The search of section~\ref{search} gives the results:
$$
\begin{array}{llcl}
C=-\frac{1}{24} & \quad B=(0,0,0) & \Rightarrow & \footnotemark f_{A,B,C}=\chi_{\{0;0\}}+2\chi_{\{0;2\}}+\chi_{\{0;4\}},\\
\\
C=\frac{1}{48} & \quad \left\{\begin{array}{c} B=\left(-\frac{1}{4},-\frac{1}{2},-\frac{3}{4}\right)\\
B=\left(-\frac{3}{4},-\frac{1}{2},-\frac{1}{4}\right) \end{array}\right. & \Rightarrow & f_{A,B,C}=2\chi_{\{1;1\}}+2\chi_{\{1;3\}},\\
\\
C=\frac{1}{24} & \quad B=\left(-\frac{1}{2},-1,-\frac{1}{2}\right) & \Rightarrow & f_{A,B,C}=2\chi_{\{2;0\}}+2\chi_{\{2;2\}}.
\end{array}
$$

\footnotetext{We have not described all $B$-values that result in $f_{A,B,C}=\chi_{\{0;0\}}+2\chi_{\{0;2\}}+\chi_{\{0;4\}}$. It turns out that the infinite family of vectors $\left(3k/2,0,-3k/2\right)$, $k\in\mathbb{Z},$ are also $B$-values in this case. This will be discussed in more detail in section~\ref{InfiniteFamilies}.}

\subsection{A general formula \label{general}}
Calculations similar to those of section~\ref{SetOfB} were carried out for the cosets $\widehat{su}(2)_1/\widehat{u}(1),\ldots,\widehat{su}(2)_{12}/\widehat{u}(1)$, from which a general pattern of suitable $B$-values emerged. In this case the $B$-values can be read directly from the matrix $A$. For a matrix $A=\mathcal{C}(A_1)\otimes \mathcal{C}(A_{k-1})^{-1},$ whose corresponding CFT is the coset model $\widehat{su}(2)_{k}/\widehat{u}(1)$, the set of $B$-values that lead to modular $f_{A,B,C}$ are
$$B=\left(\begin{array}{c}
0\\
0\\
\vdots\\
0\\
\end{array}\right) \qquad \mbox{and} \qquad
B=\underbrace{-\mathcal{C}(A_{k-1})^{-1}}_{\substack{\mbox{each column}\\\mbox{is a $B$-value}}}.
$$

\subsection{Is this the complete set of $B$-values?} \label{complete}
Whether section~\ref{general} describes all $B$-values for a matrix $A=\mathcal{C}(A_1)\otimes\mathcal{C}(A_{k-1})^{-1}$ seems to depend on whether $k$ is odd or even. For odd $k$, the search of section~\ref{search} has found no $B$-values in addition to those described in section~\ref{general}. This is not the case for even $k$, where there exist a number of additional $B$-values that lead to modular $f_{A,B,C}$. The structure of the linear combination in~(\ref{important}) seems to be a clue to the number of suitable $B$-values we can expect to find. Notice that for odd $k$ none of the linear combinations in~(\ref{important}) has a common divisor $2$ among the character multiplicities. For even $k$ many of these linear combinations can be simplified by dividing the entire combination by a common integer divisor. In such cases there seem to exist new $B$-values corresponding to each new linear combinations. This is seen more clearly in the example of section~\ref{interesting}. As in section~\ref{cosetfABC}, these conclusions are drawn based on carefully studying a number of examples. While the patterns are very interesting, we don't currently have a good understanding of why such patterns arise. This is certainly worthy of further study and we hope to undertake such work in the near future.

\subsection{An interesting example: the coset $\widehat{su}(2)_4/\widehat{u}(1)$} \label{interesting}
In example~\ref{SetOfB} we saw that for certain choices of $B$ the series $f_{A,B,C}$ was equal to the following linear combinations of coset characters:
$$\chi_{\{0;0\}}+2\chi_{\{0;2\}}+ \chi_{\{0;4\}} \qquad \mbox{or} \qquad 2\chi_{\{1;1\}}+2\chi_{\{1;3\}} \qquad \mbox{or} \qquad 2\chi_{\{2;0\}} + 2\chi_{\{2;2\}}.$$
Here the second and third linear combinations can be divided by $2$ to give
$$\chi_{\{1;1\}} + \chi_{\{1;3\}} \qquad \quad \mbox{or} \quad \qquad \chi_{\{2;0\}} + \chi_{\{2;2\}}.$$
Again, there exist values of $B$ and $C$ for which $f_{A,B,C}$ is equal to these combinations.
$$
\begin{array}{llcl}
C=\frac{1}{48} & \quad \left\{\begin{array}{c} B=\left(\frac{1}{4},\frac{1}{2},-\frac{1}{4}\right) \\B=\left(-\frac{1}{4},\frac{1}{2},\frac{1}{4}\right) \end{array}\right. & \Rightarrow & f_{A,B,C}=\chi_{\{1;1\}}+\chi_{\{1;3\}},\\
\\
C=\frac{1}{24} & \quad \left\{\begin{array}{c} B=\left(\frac{1}{2},0,-\frac{1}{2}\right) \\B=\left(-\frac{1}{2},0,\frac{1}{2}\right) \end{array}\right. & \Rightarrow & \footnotemark f_{A,B,C}=\chi_{\{2;0\}}+\chi_{\{2;2\}}.
\end{array}
$$
This phenomenon is not unique to the example $\widehat{su}(2)_4/\widehat{u}(1)$, but seems to occur for the matrix $\widehat{su}(2)_k/\widehat{u}(1)$ whenever $k$ is even.

\footnotetext{We have not mentioned all $B$-values that result in $f_{A,B,C}=\chi_{\{2;0\}}+\chi_{\{2;2\}}$. It turns out that this equation is satisfied by the infinite family of $B$-values $\left(\frac{3k+1}{2},0,\frac{-3k-1}{2}\right)$, $k\in\mathbb{Z}$. This is discussed in section~\ref{InfiniteFamilies}.}

\subsection{Infinite Families} \label{InfiniteFamilies}
For $A=\mathcal{C}(A_1)\otimes\mathcal{C}(A_3)^{-1}$ there exist infinite families of $(B,C)$--values all resulting in the same $f_{A,B,C}$. We describe two such families and give an explicit formula for the series $f_{A,B,C}$ in each case. This shows that all such $(B,C)$--combinations lead to the same value of $f_{A,B,C}$, and also proves that $f_{A,B,C}$ is modular. In this case $A$ is given by
$$A=\left(\begin{array}{ccc} \frac{3}{2} & 1 & \frac{1}{2}\\ 1 & 2 & 1\\ \frac{1}{2} & 1 & \frac{3}{2} \end{array}\right).$$ In a recent talk~\cite{SZ}, Sander Zwegers stated that a matrix $A'$ of rank $2$ could be extended to a matrix $A$ of rank $3$, and $B$ and $C$ chosen appropriately so that $f_{A,B,C}=f_{A',B',C'}$. Using his method, the $3\times3$ matrix $A$ above is an extension of the $2\times2$ matrix $$A'=\left(\begin{array}{cc}  \frac{3}{2} & -\frac{1}{2}\\ -\frac{1}{2} & \frac{3}{2} \end{array}\right).$$ $f_{A',B',C'}$ for this matrix is given explicitly in~\cite{DZ}, and it follows that, for $k\in\mathbb{Z}$
\begin{eqnarray*}
f_{A,\left(\frac{3k}{2},0,-\frac{3k}{2}\right),C} = \frac{1}{\eta(q)}\sum_{n\in\mathbb{Z}}q^{\frac{3}{4}n^2} =\chi_{\{0;0\}}+2\chi_{\{0;2\}}+\chi_{\{0;4\}},\\
f_{A,\left(\frac{3k+1}{2},0,\frac{-3k-1}{2}\right),C} = \frac{1}{\eta(q)}\sum_{n\in\mathbb{Z}}q^{\frac{3}{4}(n+\frac{1}{3})^2} =\chi_{\{2;0\}}+\chi_{\{2;2\}}.
\end{eqnarray*}
It is entirely possible that other cosets from the family $\widehat{su}(2)_k/\widehat{u}(1)$ also admit such infinite families. The searches undertaken in the current paper have not found further infinite families, but there is no reason to suggest they don't exist for larger values of $k$.

\subsection{Duality}
Define $A^*=A^{-1}$. If there exist $B\in\mathbb{Q}^r$ and $C\in\mathbb{Q}$ such that $f_{A,B,C}$ is modular, there should also exist $B^*\in\mathbb{Q}^r$ and $C^*\in\mathbb{Q}$ such that $f_{A^*,B^*,C^*}$ is modular, see~\cite{DZ}. Having determined which $(B,C)$ make $f_{A,B,C}$ modular, the corresponding $(B^*,C^*)$ are given by
$$A^* = A^{-1},\qquad \quad B^* = A^{-1}B, \qquad \quad C^* = \frac{1}{2}B^tA^{-1}B -\frac{r}{24}-C.$$

\noindent
{\it Example of using duality - the coset $\widehat{su}(k+1)_{2}/\widehat{u}(1)^k$\\}
In section~\ref{general} $B$-values were calculated for the coset $\widehat{su}(2)_{k}/\widehat{u}(1)$. The inverse of the matrix $A$ for this coset is $$A^*=A^{-1}=\mathcal{C}(A_1)^{-1}\otimes\mathcal{C}(A_k).$$
This is the matrix of the model described by the pair $\left(A_k,A_1\right)$. (More precisely it is the matrix of this model up to a permutation which doesn't effect solutions of the equations~(\ref{xeqns}) of the model.) The corresponding $B$-values are given by
$$B^* = A^{-1}B = \left(\mathcal{C}(A_1)^{-1}\otimes\mathcal{C}(A_k)\right)\left(-\mathcal{C}(A_k)^{-1}\right) = -\mathcal{C}(A_1)^{-1}\otimes I_k =-\frac{1}{2}\otimes I_k,$$
where $I$ denotes the identity matrix. Hence, the coset $\widehat{su}(k+1)_{2}/\widehat{u}(1)^k$ has $B$-values
$$B^*=\left(\begin{array}{c}
0\\
0\\
\vdots\\
0\\
\end{array}\right) \qquad \mbox{and} \qquad
B^*=\underbrace{-\frac{1}{2}\otimes I_{k}}_{\substack{\mbox{each column}\\ \mbox{is a $B$-value}}}.
$$

\section{Conclusions}
This exploratory paper was motivated by the question of when a $q$-hypergeometric series is modular. A specific $r$-fold $q$-hypergeometric series $f_{A,B,C}$ was chosen and the following question asked. Given a matrix $A$, how can $B$ and $C$ (if they exist) be calculated so that $f_{A,B,C}$ is modular? In the past, much work was done to suggest constraints on the matrix $A$, but little was known about the structure of suitable $B$-values. The aim of this paper was to better understand the $B$-values, in particular to investigate whether they followed certain patterns. As a first step all calculations were carried out for two particular families of the matrix $A$, namely $A=\mathcal{C}(A_1)\otimes\mathcal{C}(T_n)^{-1}$ and $A=\mathcal{C}(A_1)\otimes\mathcal{C}(A_{k-1})^{-1}$. Both families were chosen for their close ties to conformal field theory. The main results of the paper were formulae for calculating $B$-values appropriate to each of these families. These results arose from a simple numerical comparison between RCFT characters and $q$-series of the form~(\ref{fABC}). They indicate a relationship between $B$-values and representation theory which needs to be explored further. In some cases, however, the canonical $B$-values related to the Lie group representations come together with infinite families of associated ones which have no explanation yet. The successful search in the case $r=1$~\cite{DZ} reduced the determination of $A,B,C$-values to a set of algebraic equations. When infinite families exist, these equations will have corresponding continuous families of solutions, so that the algebraic equations alone cannot suffice for finding $B$-values and have to be supplemented by integrality conditions. The present calculations have also brought to light interesting relationships between coset characters and the series~(\ref{fABC}), see sections~\ref{cosetfABC} and~\ref{complete}. We plan to study a wider selection of cosets and matrices $A$ to see whether these relationships hold in general. If they do, it will be exciting to try to understand these patterns and to explain them in the context of conformal field theory. Although the question of when a $q$-series is modular can be viewed purely as a number theoretical problem, this paper has shown that new light can be shed on the problem by approaching it from the point of view of conformal field theory. An interdisciplinary effort in the future may be a good strategy in the hunt for a better understanding of the overlap between $q$-series and modular functions.

\appendix
\section{Statement of Nahm's conjecture and some relevant definitions} \label{Nahm}
\begin{definition}[The Bloch group]
This is the definition given in~\cite{DZ}. Let $F$ be a field. Consider the abelian group of formal sums $[z_1]+\ldots +[z_n]$, with $z_1,\ldots,z_n\in F-\{0,1\}$, satisfying
\begin{equation}
\sum_{i=1}^n(z_i)\wedge (1-z_i) = 0. \label{Bloch1}
\end{equation}
For all $x,y\in F-\{0,1\}$ with $xy\neq 1$, this group contains the elements
\begin{equation}
[x]+[1-x]\ , \qquad [x]+\left[\frac{1}{x}\right]\ , \qquad [x]+[y]+\left[\frac{1-x}{1-xy}\right]+\left[\frac{1-y}{1-xy}\right]+[1-xy]. \label{Bloch2}
\end{equation}
The Bloch group is defined as
$$\mathcal{B}(F) = \left\{[z_1]+\ldots+[z_n] \mbox{ satisfying }~(\ref{Bloch1})\right\}/\left(\mbox{subgroup generated by the elements }~(\ref{Bloch2})\right).$$
\end{definition}

\begin{definition}[Modular function]
A function which is invariant under $z\rightarrow\frac{az+b}{cz+d}$ for all $\left(\begin{array}{cc} a&b\\c&d\end{array}\right)$ belonging to some subgroup of finite index of $SL(2,\mathbb{Z}).$
\end{definition}

\begin{conjecture}[Nahm's conjecture]\label{NC}
Let $A=\left(A_{ij}\right)$ by a positive definite symmetric $r\times r$ matrix with rational entries, $B\in\mathbb{Q}^r$, and $C\in\mathbb{Q}$. Define an $r$-fold $q$-hypergeometric series $f_{A,B,C}$ by
\begin{equation}
f_{A,B,C}(q)=\sum_{n=\left(n_1,\ldots,n_r\right)\in\left(\mathbb{Z}_{\geq0}\right)^r} \frac{q^{\frac{1}{2}n^tAn+B^tn+C}}{\left(q\right)_{n_1}\ldots\left(q\right)_{n_r}}, \label{fABC2}
\end{equation}
where $(q)_n=\prod_{j=1}^n\left(1-q^j\right).$ For $i=1,\ldots,r$ we can consider the system
\begin{equation}x_i=\prod_{j=1}^r\left(1-x_j\right)^{A_{ij}}, \label{xeqns}\end{equation}
of $r$ equations in $r$ unknowns. For a solution $x=\left(x_1,\ldots,x_r\right)$ of~(\ref{xeqns}), define $\xi_x=[x_1] + \ldots + [x_r]\in\mathbb{Z}(F)$, where $F$ is the number field $\mathbb{Q}(x_1,\ldots,x_r)$. Then there exist $B\in\mathbb{Q}^r$ and $C\in\mathbb{Q}$ such that $f_{A,B,C}(q)$ is a modular function, if and only if $\xi_x$ is a torsion element of the Bloch group $\mathcal{B}(F)$ for all solutions $x=\left(x_1,\ldots,x_r\right)$ of ~(\ref{xeqns}).
\end{conjecture}

More precisely, consider the system $u_i = \sum_{j=1}^r A_{ij}v_j$, where $\exp(u_i) + \exp(v_i) = 1$, and $u_i,v_i\in\mathbb{C}$ for $i=1,2,\ldots,r$, which is essentially the logarithm of~(\ref{xeqns}). For a more detailed discussion of the solutions of~(\ref{xeqns}) see~\cite{DZ,WN}. Nahm's conjecture is correct for the case when $r=1$, see~\cite{DZ}. For $r=2$ and $r=3$ extensive computer searches have been carried out, both by Michael Terhoeven~\cite{MT} and Don Zagier~\cite{DZ}. These searches confirmed the conjecture for hundreds of cases. In one direction (the `if' in conjecture~\ref{NC}) there exists no counterexample, but in the other direction (the `only if') Masha Vlasenko and Sander Zwegers have recently found a number of counterexamples~\cite{MV}, a single one for rank $2$, and certain families for larger rank. Clearly a deeper understanding is necessary and may result from ongoing work. Any new mathematical treatment should take into account~\cite{CZ}. From a different point of view, Huang and Lee~\cite{HL} compile a complete list of positive definite, symmetric $2\times 2$ matrices with integer entries such that all complex solutions to the equations~(\ref{xeqns}) are real, and explain how this could be a useful approach to studying Nahm's conjecture.

\section{Asymptotic formulae} \label{appAsym}
The first of the asymptotic formulae used in this paper can be found in Michael Terhoeven's PhD thesis~\cite{MT}, and is a formula for computing $C$ given a symmetric matrix $A$ of rank $r$ and a vector $B=\left(b_1,\ldots,b_r\right)$:
\begin{align} \label{asymptotic1}
C &=\sum_{i=1}^r \left( \frac{\phi_2(b_i)}{2}\frac{x_i}{1-x_i} + \frac{1}{2}\phi_1(b_i)\frac{x_i}{(1-x_i)^2}F_{ii} + \frac{1}{8}\frac{x_i(1+x_i)}{(1-x_i)^3}F_{ii}^2\right) \nonumber\\
&+\sum_{i,j=1}^r\left( - \frac{1}{2}\phi_1(b_i)\frac{x_i}{1-x_i}F_{ii}\phi_1(b_j)\frac{x_j}{1-x_j} -\frac{1}{2}\frac{x_i}{(1-x_i)^2}F_{ii}F_{ij}\phi_1(b_j)\frac{x_j}{1-x_j}\right.\nonumber\\
& \left.-\frac{1}{12}\frac{x_i}{(1-x_i)^2}F_{ij}^3\frac{x_j}{(1-x_j)^2} - \frac{1}{8} \frac{x_i}{(1-x_i)^2}F_{ii}F_{ij}F_{jj} \frac{x_j}{(1-x_j)^2}\right).
\end{align}
Here $x=\left(x_1,x_2,\ldots,x_r\right)$ denotes the unique solution of the set of equations~(\ref{xeqns}), that satisfies $0<x_i<1$ for $i=1,2,\ldots,r$. The matrix $F$ is defined as $$F_{ij}=\left(\left(A^{-1}\right)_{ij}+\delta_{ij}\frac{x_i}{1-x_i}\right)^{-1},$$ and $\phi_i$ denote Bernoulli polynomial, so that
$$\phi_1(x)=x-\frac{1}{2},\qquad \phi_2(x)=x^2-x+\frac{1}{6}, \qquad \phi_3(x)=x^3-\frac{3}{2}x^2+\frac{1}{2}x.$$
Using Michael Terhoeven's method~\cite{MT,MT1}, we derive a second asymptotic formula:
\begin{align} \label{asymptotic2}
&0= \sum_{i=1}^r \left(-\frac{x_i}{6(1-x_i)^2}\phi_3(b_i) -\frac{1}{2}F_{ii}\frac{\phi_2(b_i)}{2}\frac{x_i(1+x_i)}{(1-x_i)^3}\right.\nonumber\\
&\left.- \frac{1}{8}F_{ii}^2\phi_1(b_i)\frac{x_i^3+4x_i^2+x_i}{(1-x_i)^4} -\frac{1}{48}F_{ii}^3\frac{x_i^4+11x_i^3+11x_i^2+x_i}{(1-w_i)^5}\right)\nonumber\\
& +\sum_{i,j=1}^r \left( \frac{1}{2}F_{ij}\phi_1(b_i)\frac{x_i}{1-x_i}\frac{\phi_2(b_j)}{2}\frac{x_j}{(1-x_j)^2} +\frac{1}{2}F_{ij}\frac{\phi_2(b_i)}{2}\frac{x_i}{(1-x_i)^2}\phi_1(b_j)\frac{x_j}{1-x_j}\right.\nonumber\\
& +\frac{1}{2}F_{ij}F_{jj}\phi_1(b_i)\frac{x_i}{1-x_i}\phi_1(b_j)\frac{x_j(1+x_j)}{(1-x_j)^3} + \frac{1}{2}F_{ij}F_{jj}\frac{\phi_2(b_i)}{2}\frac{x_i}{(1-x_i)^2}\frac{x_j}{(1-x_j)^2}\nonumber\\
& + \frac{1}{8}F_{ij}F_{jj}^2 \phi_1(b_i)\frac{x_i}{1-x_i}\frac{x_j^3+4x_j^2+x_j}{(1-x_j)^4} + \frac{1}{4}F_{ij}^2 \phi_1(b_i)\frac{x_i}{(1-x_i)^2}\phi_1(b_j)\frac{x_j}{(1-x_j)^2}\nonumber\\
& +\frac{1}{4}F_{ij}^2F_{jj}\phi_1(b_i)\frac{x_i}{(1-x_i)^2}\frac{x_j(1+x_j)}{(1-x_j)^3} + \frac{1}{8}F_{ii}F_{ij}F_{jj} \frac{x_i}{(1-x_i)^2}\phi_1(b_j)\frac{x_j(1+x_j)}{(1-x_j)^3}\nonumber\\
& +\frac{1}{8}F_{ii}F_{ij}F_{jj}\phi_1(b_i)\frac{x_i(1+x_i)}{(1-x_i)^3}\frac{x_j}{(1-x_j)^2} + \frac{1}{12}F_{ij}^3\frac{x_i}{(1-x_i)^2}\phi_1(b_j)\frac{x_j(1+x_j)}{(1-x_j)^3}\nonumber\\
& +\frac{1}{12}F_{ij}^3\phi_1(b_i)\frac{x_i(1+x_i)}{(1-x_i)^3}\frac{x_j}{(1-x_j)^2} + \frac{1}{16}F_{ii}F_{ij}F_{jj}^2\frac{x_i}{(1-x_i)^2}\frac{x_j^3+4x_j^2+x_j}{(1-x_j)^4}\nonumber\\
& +\frac{1}{12}F_{ij}^3F_{jj}\frac{x_i}{(1-x_i)^2}\frac{x_j^3+4x_j^2+x_j}{(1-x_j)^4} + \frac{1}{48}F_{ij}^4 \frac{x_i(1+x_i)}{(1-x_i)^3} \frac{x_j(1+x_j)}{(1-x_j)^3}\nonumber\\
& \left.+ \frac{1}{16}F_{ii}F_{ij}^2F_{jj}\frac{x_i(1+x_i)}{(1-x_i)^3}\frac{x_j(1+x_j)}{(1-x_j)^3} \right)+\sum_{i,j,k=1}^r \left(- \frac{1}{2}F_{ij}F_{jk}\phi_1(b_i)\frac{x_i}{1-x_i}\phi_1(b_j)\frac{x_j}{(1-x_j)^2}\phi_1(b_k)\frac{x_k}{1-x_k}\right.\nonumber\\
& -\frac{1}{4}F_{ij}F_{jj}F_{jk}\phi_1(b_i)\frac{x_i}{1-x_i}\frac{x_j(1+x_j)}{(1-x_j)^3}\phi_1(b_k)\frac{x_k}{1-x_k} -\frac{1}{4}F_{ii}F_{ij}F_{jk}^2\frac{x_i}{(1-x_i)^2}\frac{x_j}{(1-x_j)^2}\phi_1(b_k)\frac{x_k}{(1-x_k)^2}\nonumber\\
& -\frac{1}{8}F_{ii}F_{ij}F_{jk}F_{kk}\frac{x_i}{(1-x_i)^2}\phi_1(b_j)\frac{x_j}{(1-x_j)^2}\frac{x_k}{(1-x_k)^2} -\frac{1}{4}F_{ij}F_{jk}F_{ik}^2\frac{x_i}{(1-x_i)^2}\phi_1(b_j)\frac{x_j}{(1-x_j)^2}\frac{x_k}{(1-x_k)^2}\nonumber\\
& -\frac{1}{8}F_{ii}F_{ij}F_{jk}^2F_{kk}\frac{x_i}{(1-x_i)^2}\frac{x_j}{(1-x_j)^2}\frac{x_k(1+x_k)}{(1-x_k)^3} -\frac{1}{16}F_{ii}F_{ij}F_{jj}F_{jk}F_{kk}\frac{x_i}{(1-x_i)^2}\frac{x_j(1+x_j)}{(1-x_j)^3}\frac{x_k}{(1-x_k)^2}\nonumber\\
& -\frac{1}{12}F_{ii}F_{ij}F_{jk}^3\frac{x_i}{(1-x_i)^2}\frac{x_j(1+x_j)}{(1-x_j)^3}\frac{x_k}{(1-x_k)^2} -\frac{1}{8}F_{ij}^2F_{jk}^2F_{ik}\frac{x_i}{(1-x_i)^2}\frac{x_j(1+x_j)}{(1-x_j)^3}\frac{x_k}{(1-x_k)^2}\nonumber\\
& -\frac{1}{8}F_{ij}F_{jj}F_{jk}F_{ik}^2\frac{x_i}{(1-x_i)^2}\frac{x_j(1+x_j)}{(1-x_j)^3}\frac{x_k}{(1-x_k)^2} -\frac{1}{2}F_{ij}F_{jk}F_{kk}\phi_1(b_i)\frac{x_i}{1-x_i}\phi_1(b_j)\frac{x_j}{(1-x_j)^2}\frac{x_k}{(1-x_k)^2}\nonumber\\
& -\frac{1}{4}F_{ij}F_{jk}^2F_{kk}\phi_1(b_i)\frac{x_i}{1-x_i}\frac{x_j}{(1-x_j)^2}\frac{x_k(1+x_k)}{(1-x_k)^3} - \frac{1}{2}F_{ij}F_{jk}^2\phi_1(b_i)\frac{x_i}{1-x_i}\frac{x_j}{(1-x_j)^2}\phi_1(b_k)\frac{x_k}{(1-x_k)^2}\nonumber\\
& \left.-\frac{1}{4}F_{ij}F_{jj}F_{jk}F_{kk}\phi_1(b_i)\frac{x_i}{1-x_i}\frac{x_j(1+x_j)}{(1-x_j)^3}\frac{x_k}{(1-x_k)^2} -\frac{1}{6}F_{ij}F_{jk}^3\phi_1(b_i)\frac{x_i}{1-x_i}\frac{x_j(1+x_j)}{(1-x_j)^3}\frac{x_k}{(1-x_k)^2}\right)\nonumber\\
& +\sum_{i,j,k,l=1}^r \left( \frac{1}{4}F_{ij}F_{jk}^2F_{kl}\phi_1(b_i)\frac{x_i}{1-x_i}\frac{x_j}{(1-x_j)^2}\frac{x_k}{(1-x_k)^2}\phi_1(b_l)\frac{x_l}{1-x_l}\right.\nonumber\\
& + \frac{1}{16}F_{ij}F_{ik}^2F_{jl}^2F_{kl}\frac{x_i}{(1-x_i)^2}\frac{x_j}{(1-x_j)^2}\frac{x_k}{(1-x_k)^2}\frac{x_l}{(1-x_l)^2}\nonumber\\
& +\frac{1}{24}F_{ij}F_{ik}F_{kl}F_{jl}F_{il}F_{jk}\frac{x_i}{(1-x_i)^2}\frac{x_j}{(1-x_j)^2}\frac{x_k}{(1-x_k)^2}\frac{x_l}{(1-x_l)^2}\nonumber\\
& + \frac{1}{16}F_{ii}F_{ij}F_{jk}^2F_{kl}F_{ll}\frac{x_i}{(1-x_i)^2}\frac{x_j}{(1-x_j)^2}\frac{x_k}{(1-x_k)^2}\frac{x_l}{(1-x_l)^2}\nonumber\\
& + \frac{1}{6}F_{ij}F_{jk}F_{jl}\phi_1(b_i)\frac{x_i}{1-x_i}\frac{x_j}{(1-x_j)^2}\phi_1(b_k)\frac{x_k}{1-x_k}\phi_1(b_l)\frac{x_l}{1-x_l}\nonumber\\
& + \frac{1}{48}F_{ii}F_{ij}F_{jk}F_{jl}F_{kk}F_{ll}\frac{x_i}{(1-x_i)^2}\frac{x_j}{(1-x_j)^2}\frac{x_k}{(1-x_k)^2}\frac{x_l}{(1-x_l)^2}\nonumber\\
& +\frac{1}{8}F_{ii}F_{ij}F_{jk}F_{jl}F_{kl}^2\frac{x_i}{(1-x_i)^2}\frac{x_j}{(1-x_j)^2}\frac{x_k}{(1-x_k)^2}\frac{x_l}{(1-x_l)^2}\nonumber\\
& +\frac{1}{8}F_{ij}F_{jk}F_{jl}F_{kk}F_{ll}\phi_1(b_i)\frac{x_i}{1-x_i}\frac{x_j}{(1-x_j)^2}\frac{x_k}{(1-x_k)^2}\frac{x_l}{(1-x_l)^2} \nonumber\\
& +\frac{1}{4}F_{ij}F_{jk}F_{jl}F_{kl}^2\phi_1(b_i)\frac{x_i}{1-x_i}\frac{x_j}{(1-x_j)^2}\frac{x_k}{(1-x_k)^2}\frac{x_l}{(1-x_l)^2} \nonumber\\
& +\frac{1}{4}F_{ij}F_{jk}^2F_{kl}F_{ll}\phi_1(b_i)\frac{x_i}{1-x_i}\frac{x_j}{(1-x_j)^2}\frac{x_k}{(1-x_k)^2}\frac{x_l}{(1-x_l)^2} \nonumber\\
& \left.+\frac{1}{4}F_{ii}F_{ij}F_{jk}F_{jl}\frac{x_i}{(1-x_i)^2}\frac{x_j}{(1-x_j)^2}\phi_1(b_k)\frac{x_k}{1-x_k}\phi_1(b_l)\frac{x_l}{1-x_l}\right).
\end{align}

\end{document}